\documentclass[aps,prc ,twocolumn, nofootinbib,superscriptaddress]{revtex4-1}

\usepackage{simpler-wick}
\usepackage{graphicx}  
\usepackage{amsmath}
\usepackage{amsfonts}
\usepackage{amsbsy}
\usepackage{bm}  %
\usepackage{color}
\usepackage{tikz}
\usepackage{xcolor}
\usepackage{hyperref}
\usepackage{float}

\hypersetup{
    colorlinks,
    linkcolor={red!50!black},
    citecolor={blue!50!black},
    urlcolor={blue!80!black}
}

\newcommand{\bea}{\begin{eqnarray}}
\newcommand{\eea}{\end{eqnarray}}
\newcommand{\beq}{\begin{equation}}
\newcommand{\eeq}{\end{equation}}
\newcommand{\be}{\begin{equation}}
\newcommand{\ee}{\end{equation}}
\newcommand{\bqa}{\begin{eqnarray}}
\newcommand{\eqa}{\end{eqnarray}}

\newcommand{\tvec}{{\bf t}}

\def\mqo2{{\!\!\!}}

\newcommand{\cL}{\mathcal{L}}

\newcommand{\Mhi}{M_\text{hi}}
\newcommand*{\Mpi}{\ensuremath{m_\pi}}

\newcommand{\prob}{p}

\begin{document}
\title{The breakdown scale of pionless effective field theory in the three-nucleon sector}

\author{Andreas Ekström}\email{andreas.ekstrom@chalmers.se}
\affiliation{Department of Physics, Chalmers University of Technology, SE-412 96, Göteborg, Sweden}
\author{Lucas Platter}\email{lplatter@utk.edu}

\affiliation{Department of Physics and Astronomy, University of Tennessee, Knoxville, Tennessee 37996, USA}
\affiliation{Physics Division, Oak Ridge National Laboratory, Oak Ridge, Tennessee 37831, USA}

\date{\today}

\begin{abstract}
We make order-by-order predictions of neutron–deuteron total cross sections up to next-to-next-to-leading order in pionless effective field theory. Using Bayesian methods, we infer a posterior distribution for the breakdown scale. The result shows a mode near 100 MeV, and a combined analysis with neutron–proton scattering further sharpens the inference, placing the mode close to the pion mass scale, consistent with the expected range of pionless effective field theory.
\end{abstract}

\smallskip
\maketitle
\section{Introduction}
Effective field theories (EFTs) have become the preferred tool for analyzing the nuclear interaction~\cite{Epelbaum:2008ga,Machleidt:2011zz,Hammer:2019poc}. By systematically expanding observables in powers of a dimensionless parameter $Q = k/\Mhi$, an EFT exploits a separation of scales between the typical momentum $k$ of the process under study and a breakdown scale $\Mhi$, beyond which short-distance physics omitted from the theory need to be resolved.

Pionless EFT contains only short-range interactions~\cite{VANKOLCK1999273,Kaplan:1998tg,KAPLAN1998329} and is particularly well suited for describing low-energy processes, since the lightest exchange particle mediating the strong nuclear interaction---the pion, with mass $\Mpi \approx 138$~MeV---is not included explicitly. Its expansion parameters are $\gamma R$ and $k R$, where $\gamma$ is the inverse two-body scattering length, $k$ is the momentum scale characteristic of the process, and $R$ is the interaction range. The breakdown scale $\Mhi$ is tied to the momentum at which the EFT loses its predictive power. In the pionless theory, this scale is naturally linked to the expansion parameters and thus to the range $R$, typically assumed to be of order $\Mpi^{-1}$. 

In a previous work~\cite{Ekstrom:2024dqr}, we used Bayesian inference and order-by-order predictions of the neutron–proton ($np$) scattering cross section up to next-to-next-to-leading order (N2LO) to quantify a posterior distribution for $\Mhi$. The posterior exhibited a mode near 190 MeV, slightly above $\Mpi$, providing the first quantitative benchmark for the breakdown scale in the two-nucleon sector.

However, while $\Mhi$ is, in principle, an intrinsic property of the EFT—reflecting the separation of scales in the system under study—its practical inference may depend on the observable and kinematic regime under consideration, as well as the power counting~\cite{Bub:2024gyz}. This raises a natural and important question: Is the inferred breakdown scale consistent across systems of increasing complexity? In this work we address this question by analyzing the three-nucleon system, where more involved dynamics emerge and the three-nucleon interaction plays a significant role. 

The three-nucleon system has been studied extensively within pionless EFT since it was realized that a three-nucleon interaction is required to renormalize observables at leading order (LO)~\cite{Bedaque:1998kg,Bedaque:1999ve}. This feature is now understood as a hallmark of Efimov physics~\cite{PhysRevC.44.2303}. At N2LO, pionless EFT includes the first operator that couples $S$- to $D$-waves; however, we neglect this term in our calculation of the total cross section. For spin-dependent observables, its inclusion would be essential. 

We calculate neutron-deuteron ($nd$) scattering cross sections up to N2LO and infer a posterior for $\Mhi$ using Bayesian methods. This allows us to assess the internal consistency of the theory by comparing the inferred breakdown scale with that obtained from $np$ scattering. This paper is organized as follows: In Sec.~\ref{sec:no-pi-eft}, we briefly introduce pionless EFT necessary for calculating the $nd$ cross section. In Sec.~\ref{sec:bayes} we describe the Bayesian analysis that we have performed to infer $\Mhi$, and we end in Sec.~\ref{sec:summary} with a summary and outlook.

\section{Pionless effective field theory for neutron-deuteron scattering}
\label{sec:no-pi-eft}
The Lagrangian for the pionless EFT and resulting three-body integral equations can be defined in several ways. Here, we will mostly follow Ref.~\cite{Margaryan:2015rzg}. We used the two-body $S$-wave pionless EFT Lagrangian up to N2LO already adjusted to Z-parameterization to capture the deuteron pole and spin-singlet S-wave virtual state pole at LO.
\begin{align}
\nonumber
    \cL &= N^\dagger\left(i \partial_0+\frac{\nabla^2}{2m}\right)N\\
    \nonumber
    &\quad+ t^\dagger\left(\Delta_t -c_{0t}\left(i\partial_0+\frac{\nabla^2}{4m}+\frac{\gamma_t^2}{m}\right) \right) t
    \\
    \nonumber
    &\quad+ s^\dagger\left(\Delta_s -c_{0s}\left(i\partial_0+\frac{\nabla^2}{4m}+\frac{\gamma_s^2}{m}\right) \right) s\\
    & + y_t \left(t_i^\dagger N^T P_i N +\rm{h.c}\right)
    + y_s \left(s_i^\dagger N^T \tilde{P}_i N +\rm{h.c}\right)~,
\end{align}
where $N$ is the nucleon field, $s$ is the two-nucleon singlet field, $t$ is the two-nucleon triplet field, $\gamma_t$ is the binding momentum of the deuteron, $\gamma_s$ is the inverse scattering length in the singlet channel, and $m$ denotes the nucleon mass.

In the two-body sector at LO, this Lagrangian gives the two-body propagator
\begin{align}
    i D_{t/s}(p_0 , {\bf p}) = \frac{i}{\gamma_{t/s}-\sqrt{\frac{\bf p^2}{4}-m p_0 -i\epsilon}}~.
\end{align}

The three-body integral equations are calculated with a sharp momentum space cutoff $\Lambda$. We introduce the notation
\begin{equation}
    A(q) \otimes B(q) = \frac{1}{2\pi^2}\int_0^\Lambda dq\,q^2 A(q)B(q)~.
\end{equation}
We obtain the $nd$ t-matrix by solving the integral equation
\begin{align}
\label{eq:inteq}
\nonumber
  &\tvec_{\nu;LS}(k,p,E) = K_{\nu;LS} (k,p,E){\bf v}\\
 \nonumber
    &+\sum_{i = 1}^{\nu} \tvec_{\nu-i;LS}(k,p,E) \odot R_i(p,E)\\
    \nonumber
    &+ \sum_{i=0}^{\nu-1} K_{\nu;LS}(q,p,E) D(E-\frac{q^2}{2m},{\bf q})\otimes t_{i;LS}(k,q,E)\\
    &+K_{0;LS}(q,p,E)D(E-\frac{q^2}{2m},{\bf q}) \otimes \tvec_{\nu;LS}(k,q,E)~,
\end{align}
where the symbol "$\odot$" denotes elementwise multiplication, $L$ is the orbital angular momentum of the $n-d$ system and $S$ denotes the total spin quantum number, and $\nu$ denotes the order of the calculation and we take $\nu=0$ as LO. The vector ${\bf v} = \begin{pmatrix}
    1\\0
\end{pmatrix}$ projects onto the correct elements in the matrix $K_{\nu;LS}$. With $\tvec_{\nu;LS}$ we denote the two-component vector that combines the amplitudes for the scattering of a two-body triplet state ($t$) and a neutron ($n$) into either neutron and triplet state, i.e., $nt \rightarrow nt$, or into a neutron and a singlet state ($s$), i.e., $nt \rightarrow ns$
\begin{align}
    \tvec_{\nu;LS} = \begin{pmatrix}
        t_{\nu:LS}^{nt\rightarrow nt}\\
        t_{\nu;LS}^{nt \rightarrow ns}
    \end{pmatrix}~.
\end{align}

We neglect $SD$-coupling, and all kernel functions are therefore diagonal in the $L$ and $S$ quantum numbers.
The structure of the kernel functions depends strongly on whether the doublet or quartet channel is considered
\begin{align}
\nonumber
K_{0;L,S = 1/2}(q,p, E) &=
-\frac{2\pi}{q p }Q_l(a)
\begin{pmatrix}
    1 & -3\\
    -3 & 1
\end{pmatrix}\\
\nonumber
&-\pi H_{LO}\delta_{L0}
\begin{pmatrix}
    1 & -1\\
    -1 & 1
\end{pmatrix}~,\\
K_{0;L,S = 3/2}(q,p, E) &= \frac{4\pi}{q p }Q_l(a)
\begin{pmatrix}
    1 & 0\\
    0 & 0
\end{pmatrix}~,
\end{align}
where $a = (q^2 +p^2 -mE-i \epsilon)/q/p$ and $Q_l(a)$ denotes the Legendre function of the second kind
\begin{equation}
Q_l(a) = \frac{1}{2}\int_{-1}^1 d x \frac{P_l(x)}{x+a}~.
\end{equation}

The next-to-leading-order (NLO) kernel function only receives a correction from a subleading three-nucleon interaction that is fitted to the same parameter as the leading order three-nucleon interaction
\begin{equation}
    K_{1;LS}(q,p,E) = -\pi H_{\rm NLO}\delta_{L0}\delta_{S 1/2}
    \begin{pmatrix}
        1 & -1\\
        -1 & 1
    \end{pmatrix}~.
\end{equation}
At N2LO, we have to introduce an energy-dependent three-nucleon interaction that requires a second experimental datum for fitting~\cite{Bedaque:2002yg}
\begin{align}
\label{eq:K2}
    K_{2;LS}(q,p,E) = -\pi \left(H_{\rm N2LO} +\frac{4}{3}(m E +\gamma_t^2) H_2\right)\delta_{L0}\delta_{S\frac{1}{2}}~.
\end{align}
Higher order corrections due to two-body physics are included through the $R$-functions used in Eq.~\eqref{eq:inteq}. At NLO
\begin{align}
    R_1(p,E) = \begin{pmatrix}
        \frac{(Z_t-1)}{2\gamma_t}\left( \gamma_t +\sqrt{\frac{3}{4}p^2 -m E-i\epsilon}\right)\\
        \frac{(Z_s-1)}{2\gamma_s}\left( \gamma_s +\sqrt{\frac{3}{4}p^2 -m E - i \epsilon}\right)
    \end{pmatrix}~,
\end{align}
and at N2LO
\begin{align}
    R_2(p,E) = \begin{pmatrix}
        \frac{(Z_t-1)^2}{2\gamma_t}\left( \gamma_t +\sqrt{\frac{3}{4}p^2 -m E-i\epsilon}\right)\\
        \frac{(Z_s-1)^2}{2\gamma_s}\left( \gamma_s +\sqrt{\frac{3}{4}p^2 -m E - i \epsilon}\right)
    \end{pmatrix}~.
\end{align}
We use the method introduced by Hetherington and Schick~\cite{PhysRev.137.B935} and solve these integral equations on a complex contour. We then use this solution one more time to obtain the amplitude on the real axis. At LO and NLO, we obtain the value of the three-nucleon interaction coefficients by evaluating the three-nucleon self-energy at LO and NLO from the triton binding energy $B_t = 8.48$~MeV   (see Ref.~\cite{Vanasse:2013sda} for more details). At N2LO, the energy-dependent three-nucleon interaction $H_2$ in Eq.~\eqref{eq:K2} is included, and we additionally fit the low-energy doublet $S$-wave phaseshift to reproduce the $nd$ scattering length $a_d = 0.65 \pm 0.04$~fm~\cite{Dilg:1971gqb}. We can use the resulting amplitudes to extract channel phaseshifts and find good agreement with previously published results~\cite{Vanasse:2013sda}. 

The renormalized channel amplitudes for $nd$ scattering $M^J_{L'S',LS}(k) \equiv Z_d \cdot t_{L'S',LS}^{J; nt\rightarrow nt}$ are obtained by multiplying with the deuteron wave function renormalization factor $Z_d = 2\gamma_t/m$.
Finally, we calculate the $nd$ cross section, at relative momentum $k$, from the spin-dependent $M$-matrix
\begin{multline}
M_{m'_1,m'_2,m_1,m_2} = \sqrt{4\pi}\sum_{J,L,L'}\sum_{S,S'}\sum_{m_S,m_{S'},m_L} \sqrt{2L+1}\\
\times C_{1,\frac{1}{2},S}^{m_1,m_2,m_S}
C_{1,\frac{1}{2},S'}^{m_1',m_2',m_{S'}}C_{L,S,S}^{0,m_S,M}
C^{m_{L'},m_{S'},M}_{L',S',J}\\
\times Y_{L'm_{L'}}(\Omega) M^J_{L'S',LS}(k)~,
\end{multline}
integrated over solid angle $\Omega$, and averaged over incoming spins and summing over outgoing spins, i.e.,
\begin{equation}
    \sigma_{nd} = \frac{1}{6}\left( \frac{m}{3\pi}\right)^2
    \int d\Omega \sum_{m_1,m_2,m'_1,m'_2}|M_{m'_1,m'_2,m_1,m_2}|^2~.
\end{equation}

\section{Bayesian analysis of the breakdown scale}
\label{sec:bayes}
We seek to infer the posterior probability density $\prob(\Mhi \mid \bm{\sigma}_{nd}, I)$ for the breakdown scale $\Mhi$ of pionless EFT conditioned on a set of predicted $nd$ total cross sections at different relative momenta $k$. The vector $\bm{\sigma}_{nd}$ contains order-by-order predictions up to N2LO evaluated at a few selected momenta. The inference is also conditioned on assumptions, denoted $I$, that are outlined below. Intuitively, the posterior for $\Mhi$ reflects how well the EFT expansion converges across the orders $\nu=0,1,2$ (corresponding to LO, NLO, and N2LO) and encodes the extent to which the observed convergence pattern is consistent with the natural scales of the EFT power counting. 

Following the point-wise method presented in~\citet{Melendez:2019izc}, that we used in our previous analysis~\cite{Ekstrom:2024dqr}, we first express the sought posterior in terms of a likelihood times a prior
\begin{equation}
\prob(\Mhi|\bm{\sigma}_{nd},I) \propto \prob(\bm{\sigma}_{nd}|\Mhi,I)\cdot \prob(\Mhi|I).
\label{eq:posterior}
\end{equation}
For the likelihood $\prob(\bm{\sigma}_{nd}|\Mhi,I)$ we assume the $\nu$-th order $nd$ cross section to formally follow a series expansion~\cite{Furnstahl:2015rha}
\begin{align}
\label{eq:cs_expansion}
\sigma^{(\nu)}_{nd}(k) & = \sigma^\text{ref}_{nd}(k)
\sum_{i=0}^{\nu} c_i(k) [Q(k)]^i~,
\end{align}
\begin{figure}[t]
\centerline{\includegraphics[width=1.0\columnwidth,angle=0,clip=true]{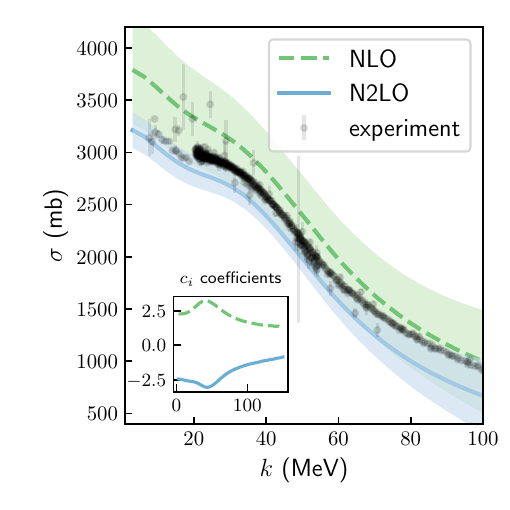}}
  \caption{\label{fig:xsecs} NLO and N2LO predictions for the $nd$ total cross section, with 68\% DoB intervals, experimental data, and the corresponding EFT expansion coefficients shown in the inset.}
\end{figure}
with a dimensionless expansion parameter $Q(k) = f(k)/\Mhi$ and EFT coefficients $c_i(k)$. We express the soft scale $f(k)$ using a sixth order soft-max function to smoothly interpolate over the low-energy scale of the deuteron binding momentum, and we assume EFT coefficients of natural size by assigning a normal prior $c_i \overset{\text{iid}}{\sim} \mathcal{N}(0, \bar{c}^2)$ with a conjugate hyperprior for the variance $\bar{c}^2 \sim \chi^{-2}(\nu_0 = 2, \tau_0^2 = 1)$. This choice places about 67\% of the probability for $\bar{c}^2$ within the natural range $ [1/3, 3]$. We take the LO prediction from pionless EFT as the reference scale $\sigma_{nd}^\text{ref}(k)$, and thus set $c_0 = 1$. In Fig.~\ref{fig:xsecs} we present NLO and N2LO predictions for the $nd$ cross sections, with  68\% degree of belief (DoB) intervals assuming $\Mhi = \Mpi$, compare with experimental values for completeness, and show the corresponding EFT coefficients $c_{i}(k)$ for $i=1,2$.

Up to N2LO, we can extract two informative coefficients $c_i(k)$. To maintain approximate independence among the expansion coefficients across momentum, we evaluate cross sections at $K = 3$ momenta 9, 61, 132 MeV. Below, we assess the robustness of our inference with respect to this momentum selection and the other assumptions. 

\begin{figure}[t]
\centerline{\includegraphics[width=1.0\columnwidth,angle=0,clip=true]{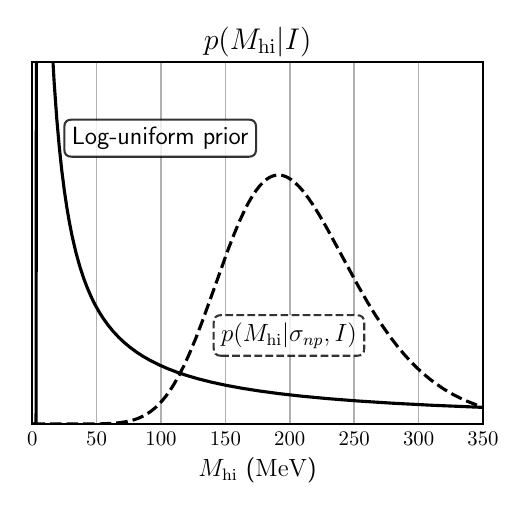}}
  \caption{\label{fig:priors} We explore two types of priors for the breakdown scale $\Mhi$ in this work. A log-uniform prior, and the posterior for the breakdown scale conditioned on the N2LO predictions for the $np$ cross sections~\cite{Ekstrom:2024dqr}}
\end{figure}

\begin{figure}[t]
\centerline{\includegraphics[width=1.0\columnwidth,angle=0,clip=true]{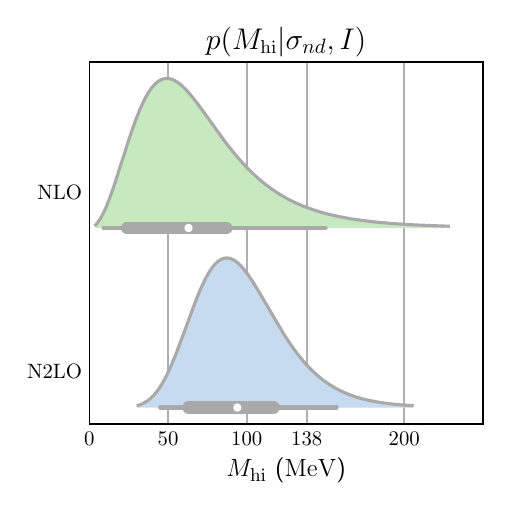}}
  \caption{\label{fig:EFT_breakdown_nd} Posteriors for the breakdown scale $\Mhi$ in pionless EFT. Thick and thin horizontal bars indicate 68\% and 95\% (highest posterior density) DoB intervals, respectively. The 68\% (95\%) DoB intervals are $[24,87]$ MeV ($[9,150]$ MeV) and $[63,117]$ MeV ($[45,157]$ MeV) at NLO and N2LO, respectively. The median values (white dots) for $\Mhi$ are 49 and 87 MeV at NLO and N2LO, respectively.}
\end{figure}

Assuming identically distributed (iid) EFT coefficients, our likelihood factorizes and the conjugate prior enables an analytical evaluation. To evaluate Eq.~\eqref{eq:posterior} it is a simple task to multiply with a prior $p(\Mhi|I)$ for the breakdown scale. To maintain similarity with our previous analysis of $np$ scattering cross sections, we first use a (scale-invariant) log-uniform distribution across a rather large interval of possible values $\Mhi \in (m_{\pi}/40,40m_{\pi})$, see solid line in Fig.~\ref{fig:priors}. 
From this, we find NLO and N2LO posteriors for the breakdown scale shown in Fig.~\ref{fig:EFT_breakdown_nd}. The posteriors mostly overlap with each other, and the N2LO mode exhibits is at a greater value, close to 100 MeV. The N2LO result is also rather robust with respect to variations of our assumptions, whereas the NLO result, for which we have fewer order-by-order predictions, is more sensitive. 

\begin{figure}[t]
\centerline{\includegraphics[width=1.0\columnwidth,angle=0,clip=true]{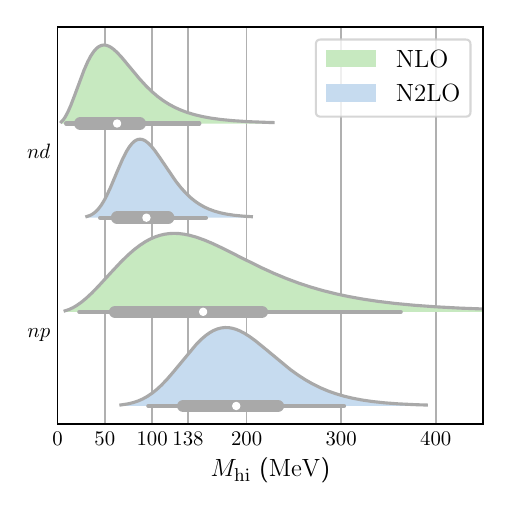}}
  \caption{\label{fig:EFT_breakdown_combined} Comparison of posteriors for the breakdown scale $\Mhi$ in pionless EFT conditioned on $nd$ predictions (top) and $np$ predictions (bottom). The $nd$ results are identical to Fig.~\ref{fig:EFT_breakdown_nd}, and the $np$ results are identical to Ref.~\cite{Ekstrom:2024dqr}.}
\end{figure}

Using a uniform prior instead of a log-uniform has a minor effect, and reasonable variation of the hyperprior for the EFT expansion coefficients also has a minor effect, at least at N2LO. Indeed, modifying the hyperprior for $\bar{c}^2$ such that $\mathbb{P}(\bar{c}^2 \leq 1) = 0.62$, by setting $\nu_0=1$ and $\tau_0=1/2$, the posteriors for $\Mhi$ are shifted to somewhat lower values. Selecting a different set of $K$ relative momenta can sligthly change the spread and mode of both the NLO and N2LO posteriors. However, the order-by-order results always remain consistent with overlapping 68\% DoB intervals, and the N2LO mode location remains in the region of 75-100 MeV. Conditioning the inference on $K>3$ relative momenta has a rather small effect on the mode locations. Using a greater number of momenta artificially shrinks the posteriors, since they are now informed by (seemingly) more order-by-order predictions. However, at some point, the finite correlation length of the EFT truncation error~\cite{Melendez:2017phj,Svensson:2023twt} will significantly violate the iid assumption underpinning our likelihood. One can extend the analysis to account for correlations using a Gaussian process, see e.g. Ref.~\cite{Millican:2024yuz} for a recent application.

The breakdown scale $\Mhi$ inferred from $nd$ data is slightly lower than that obtained from $np$ predictions, and at N2LO the 68\% DoB intervals do not overlap, see Fig.~\ref{fig:EFT_breakdown_combined}. The overlap coefficients, defined for any two probability densities $p_A(x)$ and $p_B(x)$ as
\begin{equation}
\eta(A,B) \equiv \int_{-\infty}^{+\infty} \mathrm{min} \left[ p_A(x),p_B(x)\right] \,dx.
\end{equation}
for the two posteriors $p(\Mhi|\sigma_{np},I)$ and $p(\Mhi|\sigma_{nd},I)$ are $\eta=0.41$ and $0.23$ at NLO and N2LO, respectively. This quantifies the decreasing overlap of the two posteriors with increasing EFT order. 

\begin{figure}[t]
\centerline{\includegraphics[width=1.0\columnwidth,angle=0]{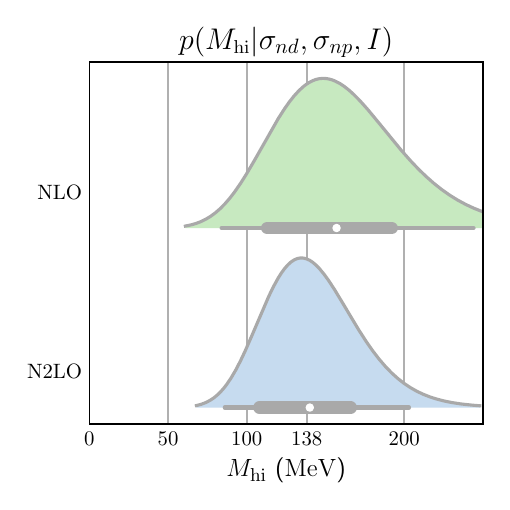}}
  \caption{\label{fig:EFT_breakdown_nd_np} Posteriors for the breakdown scale $\Mhi$ in pionless EFT conditioned $nd$ and $np$ cross section. Thick and thin horizontal bars indicate 68\% and 95\% (highest posterior density) DoB intervals, respectively. The 68\% (95\%) DoB intervals are $[113,192]$ MeV ($[84,244]$ MeV) and $[108,166]$ MeV ($[86,203]$ MeV) at NLO and N2LO, respectively. The median values (white dots) for $\Mhi$ are 149 and 135 MeV at NLO and N2LO, respectively.}
\end{figure}

We also explored the consequences of using the posterior from~\cite{Ekstrom:2024dqr}, i.e. $\prob(M_\text{hi}|\sigma_{np},I)$, as the prior in the present work, where we condition on $nd$ scattering cross sections. The posterior $\prob(M_\text{hi}|\sigma_{np},I)$ from Ref.~\cite{Ekstrom:2024dqr} is obtained using a log-uniform prior for the breakdown scale. However, the resulting posterior is virtually unchanged when using a uniform prior. Note that in the case of a uniform prior, a posterior is directly proportional to the likelihood. Thus, placing a $\prob(M_\text{hi}|\sigma_{np},I)$ prior will in our case be almost equivalent to a sequential Bayesian update of independent $\sigma_{nd}$ and $\sigma_{np}$ predictions. Indeed, 
\begin{align*}
{}& \prob(\Mhi|\bm{\sigma}_{nd},\bm{\sigma}_{np},I) \propto \prob(\bm{\sigma}_{nd},\bm{\sigma}_{np}|\Mhi,I) \prob(\Mhi|I) = \\
{}& \prob(\bm{\sigma}_{nd}|\Mhi,I) \prob(\bm{\sigma}_{np}|\Mhi,I) \prob(\Mhi|I)\propto\\
{}& \prob(\bm{\sigma}_{nd}|\Mhi,I) \prob(\Mhi|\bm{\sigma}_{np},I),
\end{align*}
where in the first equality we assume conditional independence on $\Mhi$, and in the last step we assumed a uniform prior for the breakdown scale inferred from $np$ predictions. In this approach, the posteriors at NLO and N2LO stabilize close the pion mass scale, see Fig.~\ref{fig:EFT_breakdown_nd_np}. 
This results from averaging two distributions on opposite sides of $\Mhi \approx \Mpi$. In three-body scattering, the two-body $t$-matrix is probed on- and off-shell, whereas in the two-body sector, our analysis primarily accessed the effective range expansion. We therefore speculate that the lower breakdown scale observed in the $nd$ cross section may be connected to the pion cut at $\Mpi/2$ which becomes relevant when off-shell behavior of the two-body $t$-matrix is probed.

\section{Summary}
\label{sec:summary}
Using Bayesian methods, we find a breakdown scale $\Mhi \approx 100$ MeV of the pionless EFT for $nd$ scattering. This value is somewhat smaller than the breakdown scale that we extracted recently in the two-nucleon sector~\cite{Ekstrom:2024dqr}. It is notable that in the two- and three-nucleon sector, the breakdown scale increases when moving from NLO to N2LO. Furthermore, the difference between the breakdown scales extracted from the two- and three-body sectors shows that the inferred breakdown scale of an EFT can depend on the process under consideration. Assuming independence, we perform a Bayesian sequential update, and find a combined posterior $p(\Mhi|\sigma_{nd},\sigma_{np},I)$ with a mode very close to the pion mass scale, at NLO as well as N2LO. Although interesting, this result might change when accounting for correlations.
In three-body scattering, off-shell properties of the two-body $t$-matrix are probed, unlike in the two-body case where mainly the effective range expansion is accessed. This may explain the slightly lower $\Mhi$ inferred from $nd$ cross sections, possibly due to sensitivity to the pion cut at $\Mpi/2$. Extending the analysis to spin-dependent two- and three-nucleon scattering observables is a natural next step.
\begin{acknowledgments}
The authors thank Jared Vanasse for useful discussions and providing some numerical results for benchmarking. In this work, we used the Python package 'gsum'~\cite{Melendez:2019izc} to generate most of the plots. This work was supported by the Swedish Research Council (Grants No.~2020-05127 and 2024-04681), the National Science Foundation (Grant Nos. PHY-2111426 and PHY-2412612), the Office of Nuclear Physics, and the  US Department of Energy (Contract No. DE-AC05-00OR22725).
\end{acknowledgments}

%

\end{document}